\documentclass[aps,prb,reprint,superscriptaddress,longbibliography]{revtex4-2}
\usepackage{epsfig}
\usepackage{gensymb}
\usepackage{dcolumn}
\usepackage{physics}
\usepackage{graphicx,graphics}
\usepackage{amsmath}
\usepackage{amssymb}
\usepackage{bm}
\usepackage{color}
\usepackage[utf8]{inputenc}
\usepackage[colorlinks,linkcolor={blue},citecolor={blue},urlcolor={blue}]{hyperref}
\usepackage{mathtools}
\usepackage{booktabs}
\usepackage{float}
\usepackage{mathdots}
\usepackage{hyperref}
\usepackage{academicons}
\usepackage{orcidlink}
\usepackage{mathrsfs}

\begin{document}
\title{Casimir interactions between two parallel graphene sheets carrying steady-state drift currents}

\author{Modi Ke\,\orcidlink{0000-0001-6664-1255}\,}
\affiliation{Department of Physics, \href{https://www.usf.edu}{University of South Florida}, Tampa, FL, 33620, USA}

\author{Dai-Nam Le\,\orcidlink{0000-0003-0756-8742}\,}
\affiliation{Department of Physics, \href{https://www.usf.edu}{University of South Florida}, Tampa, FL, 33620, USA}

\author{Lilia M. Woods \orcidlink{0000-0002-9872-1847}\,}
\email{lmwoods@usf.edu}
\affiliation{Department of Physics, \href{https://www.usf.edu}{University of South Florida}, Tampa, FL, 33620, USA}


\setcitestyle{numbers}
\begin{abstract}
We investigate the fluctuation–induced Casimir interactions between two parallel graphene sheets carrying steady-state drift currents. The graphene properties are modeled based on  the shifted Fermi disk model to capture the non-equilibrium optical response of the system. We find that the drift current introduces a repulsive correction to the perpendicular to the layers Casimir interaction, thereby reducing the overall attractive force.  Although the correction is repulsive, it does not overcome the underlying attraction between the layers.  It also generates a lateral force that opposes the carrier flow direction. Both contributions are studied in terms of distance and drift velocity functionalities showing pathways for Casimir force control. 

\vspace{0.2cm}

\end{abstract}

\maketitle

\section{Introduction}

Quantum vacuum fluctuations of  electromagnetic fields give rise to ubiquitous coupling between objects regardless of their properties. The associated Casimir force~\cite{Casimir1948}, in which the finite speed of light $c$ must be taken in the electromagnetic exchange, has been of great interest. It probes the interplay between the fundamental properties and geometry of the interaction objects, but it is also important for nano and micromachines for their sticktion and adhesion phenomena \cite{Woods2016,Klimchitskaya}. The discoveries of new materials have stimulated the Casimir field since properties of novel Dirac materials, topological insulators, and quasi-one dimensional systems have led to a diverse set of scaling laws, quantization effects, and even repulsion~\cite{Woods2017,Rodriguez2014,Nefedov2021}. Casimir interactions under non-equilibrium conditions have also been of much interest recently. For example, holding the objects at different temperatures or moving one object with respect to the other have been studied by different groups \cite{Bimonte2017,Antezza2008,Volokitin2011,Tomassone1997,Intravaia2014}. In addition, theoretical studies have shown that when two charge-neutral objects move parallel to each other, quantum electromagnetic fluctuations would produce a drag force even in vacuum~\cite{Milton2016,Milton2023}.

Casimir phenomena are also influenced by having drift charge carriers, for example in the case of bias voltage resulting in dc currents in the objects. Drift carriers affect the electromagnetic waves inside each material, and they also couple to the vacuum fluctuations exchanged between the objects. The case of semiconducting plates supporting dc currents have been considered where the dielectric response was taken via the semi-classical Boltzmann equation \cite{Dalvit2008} or the Drude model
\cite{Shapiro2010}. The Casimir-Polder interaction between a nanoparticle and a current-carrying substrate was also studied in \cite{Shapiro}, where one distinguishes between two models for the dielectric response of the substrate. In one model it is assumed that the fluctuations arise predominantly from the  lattice of the material taken to be in equilibrium, while the other model takes the electron plasma approximation with Doppler–shifted noise as the dominant source with an idealized separation of contributions from lattice and electrons~\cite{Shapiro}. 
 
In this paper, we examine the Casimir interaction between two dc current carrying parallel graphene sheets. As found in the previous papers \cite{Volokitin_2013,Farias}, stronger effects are found in systems with larger currents. Graphene can support carriers with very large drift velocities~\cite{Yamoah2017,Yin2014,Dorgan2010}, thus this effect may be stronger in interacting graphene sheets.
For this purpose, we resolve the boundary condition of two parallel graphene sheets carrying a drift current. Then the Maxwell stress tensor is used to find the fluctuation induced forces in terms of the reflection coefficients of the two layers. 

Similar setup has been considered by~\cite{Volokitin_2013} where the effect of the drift current is treated simply as a Doppler shift. Instead in our paper we utilize a phenomenological shifted Fermi disk (SFD) model for the conductivity, which more accurately captures the dielectric response of the current carrying graphene and reduces to a Doppler-like transformation only for two dimensional electron gas~\cite{Sabbaghi1,Sabbaghi2}. 

Unlike earlier approaches that represent the effect of a drift current as a simple Doppler shift in the electromagnetic response or through semiclassical Drude–Boltzmann transport~\cite{Shapiro,Shapiro2010,Volokitin_2013}, the SFD model also captures the intrinsic redistribution of carriers in momentum space caused by steady-state drift. This microscopic description naturally accounts for the graphene’s linear Dirac dispersion and the resulting anisotropic conductivity tensor, thereby going beyond the Doppler-like approximations valid only for parabolic two-dimensional electron gases~\cite{Sabbaghi1,Sabbaghi2}. The SFD model thus provides a physically consistent framework for treating non-equilibrium optical response in current-carrying graphene.

\section{Graphene layers and current induced modifications in the optical response}

\begin{figure}[h!]  
    \centering
    \begin{minipage}{0.4\textwidth}
        \centering
    \includegraphics[width=0.7\textwidth]{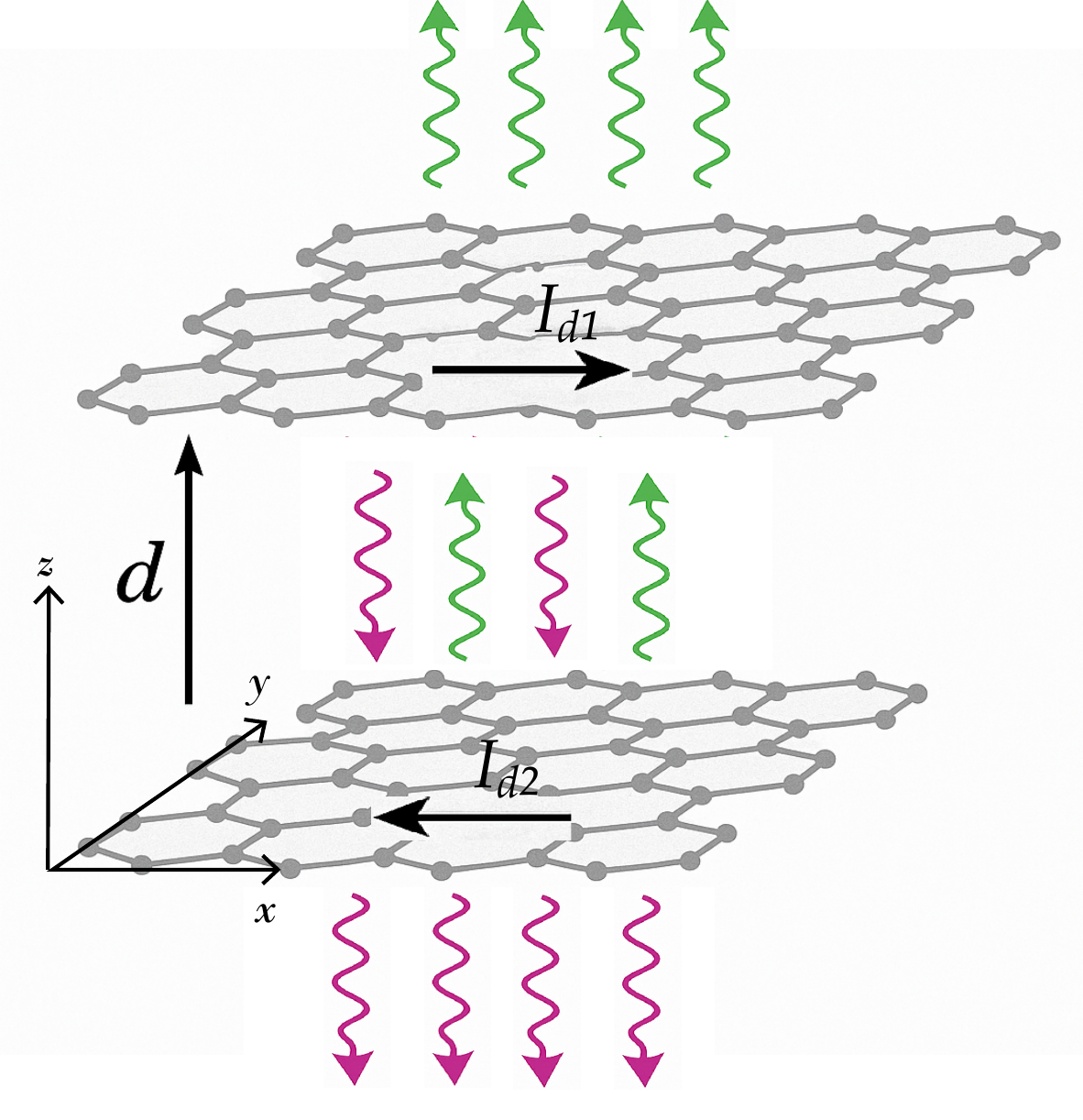}
        \caption{System setup: two parallel graphene sheets separated by a distance $d$ with drift currents  $I_{d,1,2}$ corresponding to electrons moving with drift velocities $\boldsymbol{v}_{d,1,2}$ respectively. }\label{fig1}
    \end{minipage}\hfill
\end{figure}

The system under consideration consists of two parallel graphene sheets separated by a distance $d$ as shown in Figure \ref{fig1}. Each graphene sheet has drift carriers giving rise to a steady-state dc currents $I_{1,2}$ driven by an external voltage bias.

An integral part of our modeling is the graphene optical response in the presence of a drift current. The preferred direction established by  the current breaks the local rotational invariance and the time reversal symmetry in the nonlocal response leading to modified plasmonic group velocities~\cite{Sabbaghi1,Sabbaghi2}. In the presence of a steady-state drift current, the electrons reach a non-equilibrium satationary state by acquiring momentum and kinetic energy from the drain-source voltage while simultaneously dissipating part of that energy through electron scattering mechanisms. 

The SFD non-equilibrium carrier distribution  
 ~\cite{Sabbaghi1,Sabbaghi2} can be written as $n_F (E,k)=n_F^0 (E) + \Delta n_F (E,k)$. Within this model in the low temperature ($k_BT \ll E_F$) and small drift velocity regime, the equilibrium Fermi distribution $n_F^0 (E) = 1/(e^{(E-E_F)/k_BT}+1)$ becomes modified by an additional contribution $\Delta n_F(E,\boldsymbol{k})\approx-E_F \delta(E-E_F)\cos{\theta_k}k_{\rm shift}/k_F$ where $\theta_k$ is the angle of the momentum $\boldsymbol{k}$ with the $x$ axis and $\delta(E-E_F)$ represent the Dirac delta function. Here $E_F=\hbar v_F k_F$ is the Fermi energy of the graphene electrons with Fermi momentum $k_F$ and Fermi velocity $v_F$, while $k_{\rm shift}=k_F v_{\rm d}/v_F$ is the momentum of the driven electrons with drift velocity $v_{\rm d}$. 

\begin{widetext}

The conductivity tensor components of graphene with a drift current can be written as 

\begin{eqnarray}\label{condtotal}
&&\sigma_{xx,yy}(\boldsymbol{q},\omega)=\sigma_0(\boldsymbol{q},\omega)+\Delta\sigma_{xx,yy}(\boldsymbol{q},\omega),
\end{eqnarray}
\begin{eqnarray}\label{condeq}
&&\sigma_0(\boldsymbol{q},\omega)= \sigma_u \left\{ \frac{\hbar \omega}{\sqrt{(\hbar \omega)^2 - (\hbar v_F q)^2}}
\left[1+G\left(\frac{\hbar\omega+2|E_F|}{\hbar v_Fq}\right)-G\left(\frac{\hbar\omega-2|E_F|}{\hbar v_Fq}\right) \right]-\frac{8i}{\pi}\frac{\hbar \omega|E_F|}{(\hbar v_F q)^2} \right\},
\end{eqnarray}

\end{widetext}
where $\sigma_0(\boldsymbol{q},\omega)$ is the graphene conductivity without a drift current, $G(x)=-\frac{1}{\pi}(x\sqrt{1-x^2}-\arccos{x})$, and $\sigma_u=e^2/4\hbar$ is the universal graphene conductivity \cite{Wunsch2006, Hwang2007, Fei2012, Fizeau}. The current induced corrections can be found as a result of $\Delta n_F(E,\boldsymbol{k})$ within the SFD model in the Kubo's formula for optical conductivity~\cite{Sabbaghi1,Sabbaghi2}.
 Using the Dirac cone model for the graphene spectrum we then arrive at the following modification to the conductivity:
%
%
%
\begin{widetext}

\begin{figure}[h!]  
    \centering
    \begin{minipage}{1\textwidth}
        \centering
        \includegraphics[width=0.8\textwidth]{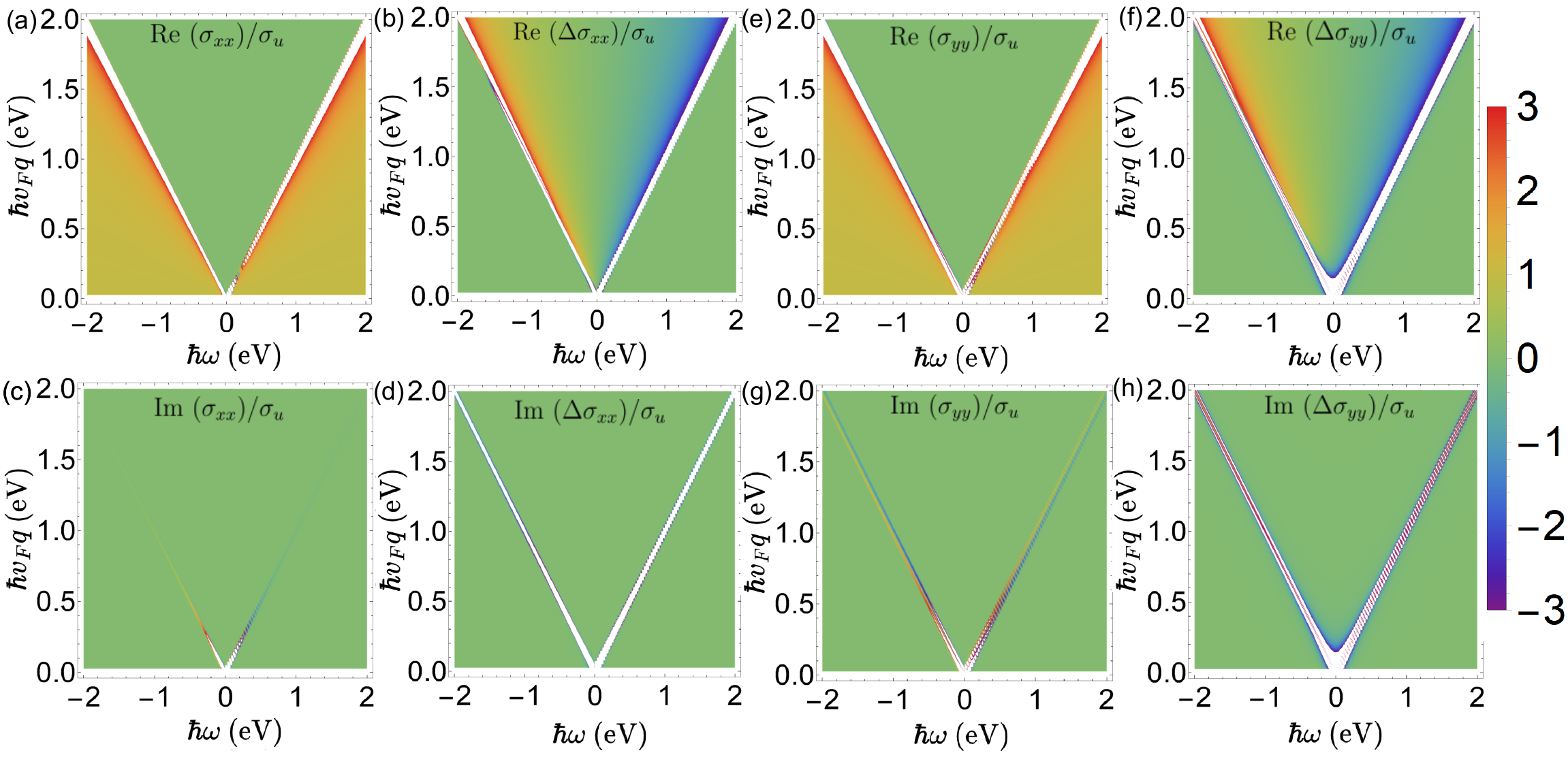}
        \caption{Color map of the optical conductivity components of graphene supporting a drift current in the $\hbar v_Fq$ vs $\hbar\omega$ space for (a) $\text{Re }(\sigma_{xx})/\sigma_u$; (b) $\text{Re }(\Delta \sigma_{xx})/\sigma_u$; (c) $\text{Im }(\sigma_{xx})/\sigma_u$; (d) $\text{Im }(\Delta \sigma_{xx})/\sigma_u$; (e) $\text{Re }(\sigma_{yy})/\sigma_u$; (f) $\text{Re }(\Delta \sigma_{yy})/\sigma_u$; (g) $\text{Im }(\sigma_{yy})/\sigma_u$; (h) $\text{Im }(\Delta \sigma_{yy})/\sigma_u$. Here we have used $\beta_d=0.1$, $E_F=0.01 {\rm eV}$ and $\theta_q=0$.}\label{fig2}
    \end{minipage}
\end{figure}

\begin{eqnarray}\label{condcorrection}
\Delta\sigma_{xx(yy)}(\boldsymbol{q},\omega) \approx \beta_d \sigma_u \sum\limits_{\xi = \pm 1} \int_0^{2\pi} d\theta_k && \left\{
\frac{2 i E_F^2 \cos{\theta_k}}{\pi^2 \left[ \left(\xi\sqrt{E_F^2+( \hbar v_F q)^2 + 2 E_F  (\hbar v_F q) \cos{(\theta_k-\theta_q)}}- E_F \right) ^2- (\hbar \omega)^2-2i \eta (\hbar \omega) \right] } \times \right. \nonumber\\
&& \left. \times \left[1\pm\xi\frac{E_F\cos{2\theta_k} + (\hbar v_F q) \cos{(\theta_k+\theta_q)}}{\sqrt{E_F^2+(\hbar v_F q)^2+2 E_F (\hbar v_F q) \cos{(\theta_k-\theta_q)}}}\right] \right\},
\end{eqnarray}
\end{widetext}
where $\xi$ represents the band indices of graphene with $+$ denoting conduction and $-$ denoting valence Dirac bands. The plus sign $+$ before $\xi$ corresponds to the $xx-$component and minus sign $-$ to the $yy-$component of the conductivity, while $\eta=1~{\rm meV}$ is the band broadening. The integration in Eq.~\eqref{condcorrection}  is performed over $\theta_k$ which is the angle of the momentum vector $\boldsymbol{k}$ has with respect to the $x$ axis. Eq.~\eqref{condcorrection} shows a linear relationship with $\beta_d=v_{\rm d}/v_F$, the ratio between the drift and Fermi velocities, and a quadratic relationship with the Fermi energy $E_F$.

In Fig.~\ref{fig2} we show density maps of the real and imaginary parts of the graphene conductivity $\sigma_{xx}=\sigma_{0}+\Delta\sigma_{xx}$, $\sigma_{yy}=\sigma_{0}+\Delta\sigma_{yy}$ and their nonequilibrium corrections $\Delta\sigma_{xx}$, $\Delta\sigma_{yy}$  (see Eq.~\eqref{condcorrection}) in the $\hbar v_F q$ vs $\hbar \omega$ space. It can be seen that the non-equilibrium correction is significant primarily near the region of  $\hbar\omega\sim\hbar v_F q$  meaning that the wave vector $q  \approx \omega/v_F\gg\omega/c$  mostly captures the role of the drift current. Fig.~\ref{fig2} shows that $\Delta\sigma_{yy}$ has a larger contribution to the total conductivity as compared to the $x$-direction, although both corrections increase in magnitude as $\beta_d$ is increased. We find that larger drift currents and larger Fermi energies induce stronger modifications in the graphene optical response.  In Ref.~\cite{Sabbaghi1} similar plot has been done for the polarization of graphene instead of its conductivity where the correction is similarly maximal around the region of  $\hbar\omega\sim\hbar v_F q$.

\section{The Maxwell stress tensor and force components}
The fluctuation induced force of the electromagnetic excitations is closely related to the Maxwell stress tensor $\stackrel{\leftrightarrow}{\mathscr{T}}$~\cite{Lifshitz,Polder}. In addition to the interaction directed along the $z-$axis, the drift current induces a  lateral Casimir force component. Considering the case of currents flowing in the $x$-direction, these are given as

\begin{widetext}
\begin{eqnarray}\label{forcez}
F_z && =[\mathscr{T}_{zz}]_{z=0^+}-[\mathscr{T}_{zz}]_{z=0^-}
\nonumber\\
&& =\frac{1}{4\pi}\int^{\infty}_0d\omega\int \frac{d^2q}{(2\pi)^2} \left\{ \left[\langle E_zE_z^*\rangle-\langle B_xB_x^*\rangle-\langle B_yB_y^*\rangle \right]_{z=0^+} - \left[\langle E_zE_z^*\rangle-\langle B_xB_x^*\rangle-\langle B_yB_y^*\rangle \right]_{z=0^-} \right\}, \\
\label{forcex}
F_x && =[\mathscr{T}_{xz}]_{z=0^+}-[\mathscr{T}_{xz}]_{z=0^-} \nonumber\\
&& =\frac{1}{4\pi}\int^{\infty}_0d\omega\int \frac{d^2q}{(2\pi)^2} \left\{ \left[\langle E_xE_z^*\rangle+\langle E_x^*E_z\rangle+\langle B_xB_z^*\rangle+\langle B_x^*B_z\rangle \right]_{z=0^+} \right. \nonumber\\
&& \quad \quad \quad \quad \quad \quad \quad \quad \quad \quad \quad \quad \quad \quad \quad  \left. - \left[ \langle{E_xE_z^*}\rangle+\langle{E_x^*E_z}\rangle+\langle B_xB_z^*\rangle+\langle B_x^*B_z\rangle \right]_{z=0^-} \right\},
\end{eqnarray}

\end{widetext}
where $E_{x,y,z}$ and $B_{x,y,z}$ are the components of the electric and magnetic fields, respectively. All $\langle{...}\rangle$ terms correspond to fluctuations of the fields, expressed through their correlation functions evaluated at $z=0_{\pm}$, i.e., as the boundary is approached from the top and bottom of the $z$-axis, respectively (see Fig.~\ref{fig1}). In alternative formulations, these correlations are often obtained from the electromagnetic Green's functions of the system~\cite{Volokitinbook}. In that approach, the fluctuating sources are encoded through the dyadic Green's tensors, which propagate the fields subject to the given boundary conditions, thereby providing a natural link between material response and fluctuating fields. By contrast, in Rytov's fluctuational electrodynamics~\cite{Lifshitz}, the starting point is the correlation functions of the fluctuating current densities inside the media. The field correlations are then derived by propagating these stochastic sources through Maxwell's equations. The Green's function and Rytov approaches are formally equivalent, but differ in emphasis: the former focuses on constructing the field propagators explicitly, while the latter emphasizes the statistical properties of the microscopic sources. In this work we adopt the Rytov framework, which is particularly convenient for treating boundary conditions at planar interfaces and for expressing fluctuation--dissipation relations directly in terms of material response functions.

To calculate explicitly the force components in Eqs.~\eqref{forcez}-\eqref{forcex}, the electromagnetic boundary conditions are resolved for the planar symmetry in Fig.~\ref{fig1}. Details can be found in Appendix \ref{appendix}. The current fluctuations in the graphene layers are taken into account via the fluctuation dissipation theorem~\cite{Lifshitz,Polder}, for which $\langle j_{f\alpha}(\boldsymbol{q},\omega)j_{f\beta}(\boldsymbol{q},\omega)\rangle=\frac{\hbar \omega}{\pi} \left(\frac{1}{2}+n_B(\omega) \right){\rm Re}[\sigma_{\alpha\beta}(\boldsymbol{q},\omega)]$ where $n_B(\omega)=1/(e^{(\hbar \omega)/k_BT}-1)$ with $T$ denoting  temperature. In the case of drift carriers flowing through each graphene layer, this expression needs to be modified since the fluctuation dissipation theorem is valid in equilibrium. Thus, the original stationary reference frame is changed via a Lorentz transformation to a frame attached to the drifting carriers, which means that the conductivity must take into account the presence of current density  with magnitude $I_d\approx n_s \beta_d v_F$ (model discussed earlier). Assuming the drift velocity is much smaller than the speed of light, one can  omit $(v_d/c)^2$ and higher corrections in the Lorentz transformation~\cite{Volokitinbook}. As a result, there is a Doppler shifted frequency,  
 $\omega'=\omega-\boldsymbol{q}\cdot \boldsymbol{v}_{d,1}$ for the top layer and $\omega''=\omega-\boldsymbol{q}\cdot \boldsymbol{v}_{d,2}$ for the bottom layer.

With the obtained reflection coefficients from the electromagnetic boundary conditions and assuming that both graphene layers have the same temperature, we find that
\begin{widetext}
\begin{eqnarray}\label{forcez2}
&&F_z=\frac{\hbar}{2\pi^3}\int^{\infty}_0d\omega\int_{q<\omega/c} d^2q \big\{\frac{-k_z}{|\Delta_p(\omega',\omega'')|^2}\bigg[\bigg(\frac{1}{2}+n(\omega'')\bigg){\rm Re}(R_{2p}(\omega')e^{2ik_zd})({\rm Re}R_{1p}(\omega'')-|R_{1p}(\omega'')|^2)\nonumber\\
&&+\bigg(\frac{1}{2}+n(\omega')\bigg){\rm Re}R_{1p}(\omega'')({\rm Re}R_{2p}(\omega')-|R_{2p}(\omega')|^2)\bigg]+(p\leftrightarrow s)\big\}\\
&&+\frac{\hbar}{2\pi^3}\int^{\infty}_0d\omega\int_{q>\omega/c} d^2q e^{-2|k_z|d}\big\{\frac{-ik_z}{|\Delta_p(\omega',\omega'')|^2}\bigg[
\frac{1+n(\omega')+n(\omega'')}{2}{\rm Im}(R_{1p}(\omega'')R_{2p}(\omega'))\nonumber\\
&&+\frac{n(\omega')-n(\omega'')}{2}{\rm Im}(R^*_{1p}(\omega'')R_{2p}(\omega'))\bigg]+(p\leftrightarrow s)\big\},\nonumber
\end{eqnarray}
\begin{eqnarray}\label{forcex2}
&&F_x=\frac{\hbar}{2\pi^3}\int^{\infty}_0d\omega\int_{q<\omega/c} d^2q q_x\big\{\frac{1}{|\Delta_p(\omega',\omega'')|^2}\bigg[(n(\omega')-n(\omega''))({\rm Re}R_{1p}(\omega'')-|R_{1p}(\omega'')|^2)({\rm Re}R_{2p}(\omega')-|R_{2p}(\omega')|^2)\nonumber\\
&&-\frac{1}{2}\bigg(\frac{1}{2}+n(\omega'')\bigg)({\rm Re}R_{1p}(\omega'')-|R_{1p}(\omega'')|^2)(|R_{2p}(\omega')-1|^2+|R_{2p}(\omega')e^{2ik_zd}-1|^2)+(p\leftrightarrow s)\big\}\\
&&+\frac{\hbar}{2\pi^3}\int^{\infty}_0d\omega\int_{q>\omega/c} d^2q q_xe^{-2|k_z|d}(n(\omega')-n(\omega''))\frac{{\rm Im}R_{1p}(\omega''){\rm Im}R_{2p}(\omega')}{|\Delta_p(\omega',\omega'')|^2}+(p\leftrightarrow s)\nonumber,
\end{eqnarray}
\end{widetext}
where $\Delta_{s(p)}(\omega',\omega'')=1-e^{2ik_zd}R_{1s(p)}(\omega'')R_{2s(p)}(\omega')$ and $k_z = \sqrt{\omega^2 / c^2 - q^2}$.  Also,  
$R_s(\omega)=\sigma_{{\rm T}}(\omega)/\left(\dfrac{c^2 k_z}{2\pi \omega} + \sigma_{{\rm T}}(\omega) \right)$ and $R_p(\omega)=\sigma_{{\rm L}}(\omega)/\left(\dfrac{\omega}{2\pi k_z} + \sigma_{{\rm L}}(\omega) \right)$ are the reflection coefficients corresponding to s (TE) and p (TM) polarizations of the EM field. For each layer, $R_s$ and $R_p$ are modified due to the presence of a drift current  with their respective Doppler shifted frequencies $\omega'$ and $\omega''$. The conductivity components in longitudinal (L) and transverse (T) directions are  $\sigma_{{\rm L}}=\sigma_{{\rm xx}}\cos ^2{\theta_q}+\sigma_{{\rm yy}}\sin ^2{\theta_q}$ and $\sigma_{{\rm T}}=\sigma_{{\rm xx}}\sin ^2{\theta_q}+\sigma_{{\rm yy}}\cos ^2{\theta_q}$. Also, $\zeta_{s(p)}(\omega)=1/(1-R_{s(p)}(\omega))$ with the Doppler shifted frequencies for each layer. 

Eqs.~\eqref{forcez2}-\eqref{forcex2} show that the vertical and lateral components of the current-modified Casimir interaction are composed of propagating ($q<\omega/c$) and evanescent ($q>\omega/c$) contributions, similar to the graphene-graphene interaction in equilibrium conditions~\cite{physics5040066}. We note that in the absence of a current, $F_x$ becomes zero as all the integrands are odd in $q_x$ leading to exact cancellations upon $q_x$ and $-q_x$ integration. We also find that in this case,  Eq.~\eqref{forcez2} can be evaluated by rotating the frequency integral into the complex plane, which transforms the $F_z$ expression into the conventional Lifhsitz-like form in the imaginary frequency domain~\cite{Woods2010,Klimchitskaya}. As a result, the Casimir force between the graphene sheets in the absence of drift currents is obtained as $F_0=-\frac{3\hbar c \alpha}{32\pi d^4}$ where $\alpha$ is the fine structure constant.

\begin{figure}
    \begin{minipage}{0.5\textwidth}
        \centering
        \includegraphics[width=0.9\textwidth]{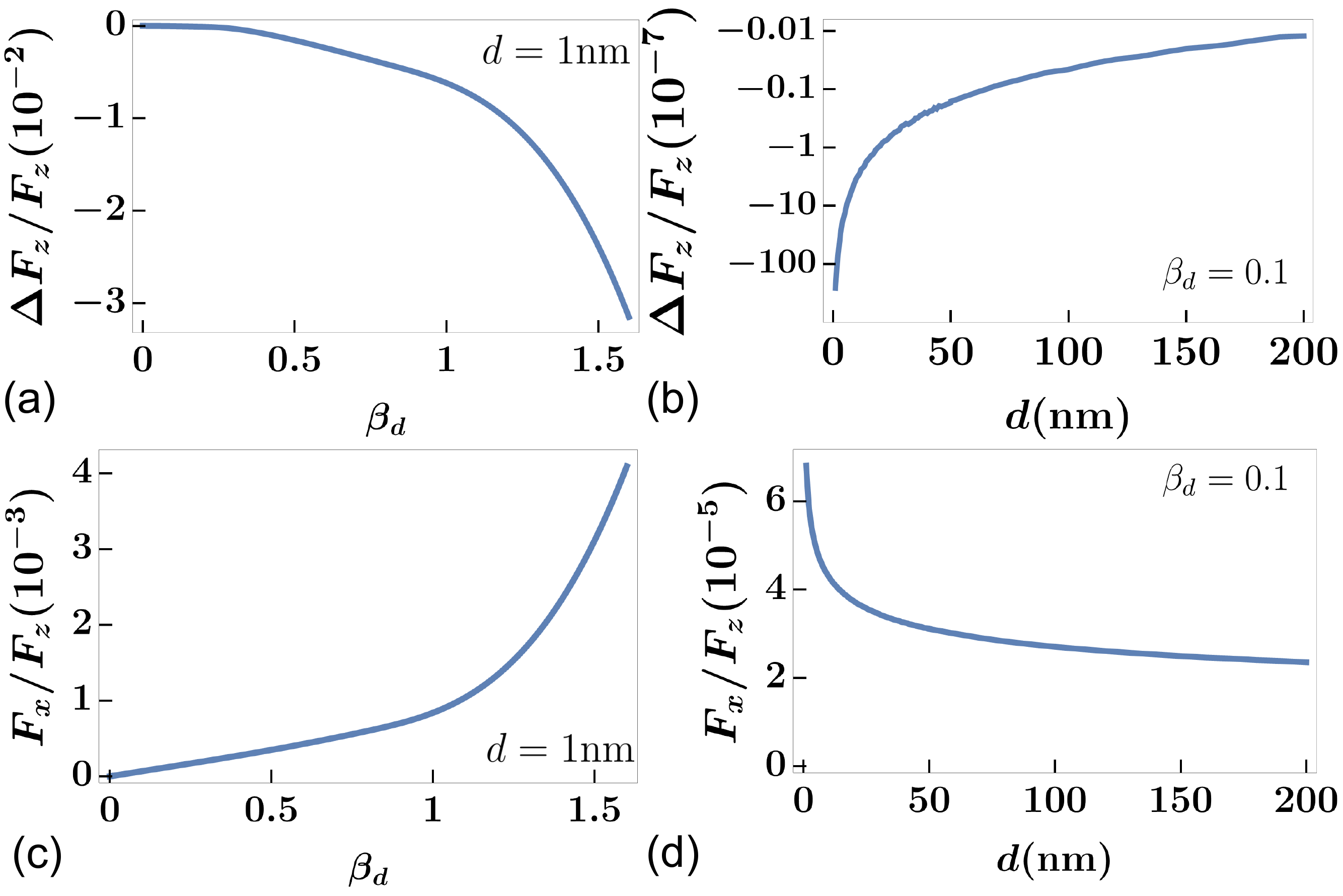}        \caption{Graphene-graphene Casimir interaction in the presence of a drift current in the top layer: (a) $\Delta F_z/F_z$  as a function of $\beta_d$ at $d=1$ nm; (b) $\Delta F_z/F_z$ as a function of $d$ for $\beta_d=0.1$; (c) $F_x/F_z$ as a function of $\beta_d$ for $d=1$ nm; (d) $F_x/F_z$ as a function of $d$ for $\beta_d=0.1$ Here $\Delta F_z= F_z- F_0$, where $F_z$ is the interaction along the $z$-axis when the drift current flows in the top graphene layer and $F_0=-\frac{3\hbar c\alpha}{32\pi d^4}$ is the equilibrium Casimir force. In all cases, the graphene Fermi level is taken as $E_F=0.1$ eV and $T=240$ K. }\label{fig5}

    \end{minipage}\hfill
\end{figure}

At this point it is important to note that the Casimir interaction in the presence of drift currents has also been studied in a system composed of a nanoparticle and a 3D substrate, which is also relevant for the results obtained in Eqs.~\eqref{forcez2}-\eqref{forcex2}.  In particular, Shapiro~\cite{Shapiro} developed a microscopic model for fluctuation-induced interactions between a nanoparticle and a current-carrying medium, identifying distinct physical pictures depending on whether the underlying fluctuations originate from the drifting electron plasma or from the lattice. In one of the considered scenarios, the contribution to the force comes
from the moving electron plasma, while the lattice dissipation is neglected; in the other case, the fluctuation of the lattice dominates and thus the moving electrons would not affect the fluctuation-induced interaction. If the primary source of fluctuations originates from the lattice, it would be necessary to consider an effective  temperature of the lattice that is different than the effective temperature of the drifting electrons, which complicates the problem as it is difficult to find these two different temperatures~\cite{Shapiro}. For the fluctuations originating primarily from the electrons a Doppler shift is induced in the fluctuation frequencies $\omega',\omega''$ due to the carrier motion.  In this work, we regard electrons as the primary source of fluctuations with a drift current in stationary medium.

Volokitin and Persson~\cite{Volokitin_2013}  
modeled the drift current in one of the two parallel graphene layers via a Lorentz transformation—formally equivalent to a current-free medium moving along the $x$-axis at  $v_d$. However, this treatment differs fundamentally from the physical situation in graphene, where the lattice remains static in the laboratory frame while only the carriers are moving. In the electron rest frame induced by the Lorentz boost, the lattice appears to move oppositely, yielding a dielectric response that deviates from the equilibrium case of static electrons and lattice. Therefore, a simple Doppler shift cannot adequately describe the graphene conductivity under drift. In~\cite{Sabbaghi1} it has been shown that the SFD model and Doppler-shift approximation coincide only for an ideal two-dimensional electron gas, with discrepancies arising in graphene due to its Dirac nature. Here, we adopt a similar framework as in~\cite{Volokitin_2013} while implementing the full SFD model to achieve a comprehensive description of the graphene optical response under drift. 

\section{Current induced modifications in the interaction}

To understand the role of drift currents in the graphene Casimir interaction, here we present results for the numerical calculations of Eqs.~\eqref{forcez2}-\eqref{forcex2}  for the case of charge carriers flowing along the $x$-axis in the top graphene layer with Doppler shifted frequency $\omega '=\omega - q_x v_{\rm d}$ (for the bottom layer there is no drift velocity). The results in Fig.~\ref{fig5} (a)-(b) show how the modified interaction along the vertical direction depends on $\beta_d$ and $d$.  At $\beta_d=0$ when there is no current, $\Delta F_z= 0$ and we recover the usual equilibrium Casimir force between two graphene sheets $F_0=-\frac{3\hbar c \alpha}{32\pi d^4}$ previously discussed in literature~\cite{Woods2017,Woods2010}. In the presence of a drift current, the Casimir force $F_z$ is modified by $\Delta F_z$ found to be with a positive sign. This repulsive graphene-graphene correction along the $z-$axis has a slightly slower than a quadratic $\beta_d^2$ dependence, as shown in Fig.~\ref{fig5} (a). We also note that  $\Delta F_z$ has a different distance dependence when compared to $F_0$; in particular, we find that  
$\Delta F_z\sim \frac{1}{d^{5.8}}$ indicating a shorter range interaction compared to the equilibrium case with its $F_0\sim\frac{1}{d^4}$ distance power law. Fig. \ref{fig5} (a, b) shows that this current-induced repulsive correction is most significant at small separations and large drift currents.

\begin{figure}
    \begin{minipage}{0.45\textwidth}
        \centering
        \includegraphics[width=0.9\textwidth]{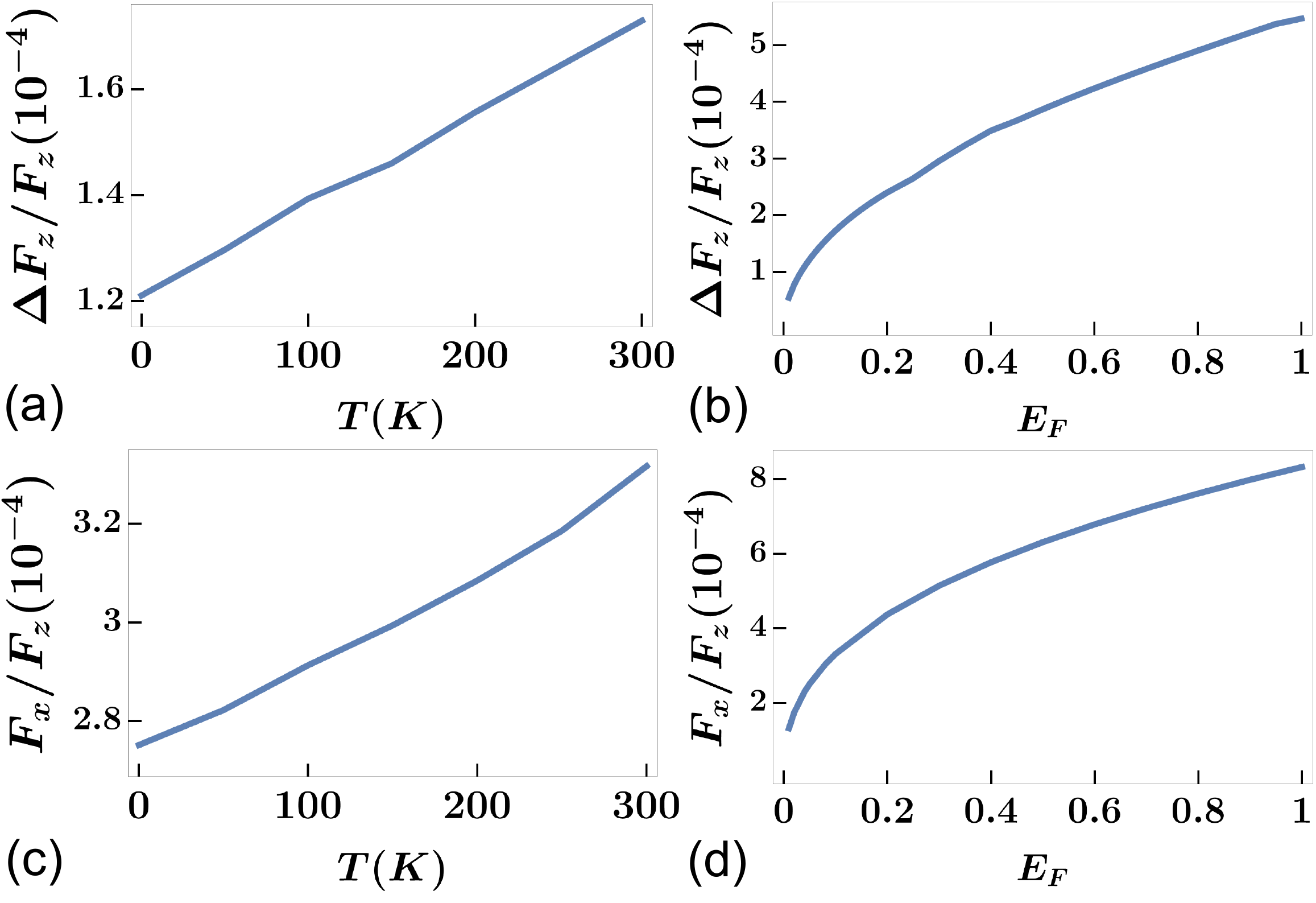}        \caption{Graphene-Graphene Casimir interaction in the presence of a drift current in the top layer: (a) $\Delta F_z/F_z$  as a function of temperature $T$; (b) $\Delta F_z/F_z$ as a function of $E_F$; (c) $F_x/F_z$ as a function of temperature $T$ ; (d) $F_x/F_z$ as a function of $E_F$. Here we used $d=10$  nm, and $\beta_d=0.5$. Also, $F_z$ is the interaction along the $z$-axis when the drift current flows in the top graphene layer and $F_0=-\frac{3\hbar c\alpha}{32\pi d^4}$ is the equilibrium Casimir force. In (a)(c), the graphene Fermi level is taken as $E_F=0.1$ eV and in (b)(d) the temperature is $T=300$ K.}\label{fig4} 

    \end{minipage}\hfill
\end{figure}

As discussed earlier, due to the drifting electrons there is also a lateral Casimir-like force $F_x$ responsible for dissipating energy in the system. The lateral force opposes the direction of the driting carriers, and it is similar to the noncontact friction arising between two graphene sheets in relative motion \cite{Shapiro,Farias,Gorbachev2012} . Fig.~\ref{fig5} (c) shows that the lateral force is several orders of magnitude smaller when compared to the equilibrium $F_0$. It has a linear dependence upon $\beta_d$ for $\beta_d<1$, which is consistent with previously reported results for small drift current~\cite{Volokitin_2013}, and it has a nonlinear behavior as $\beta_d$ becomes larger.  In Fig.~\ref{fig5} (d) we also show how $F_x$ changes as a function of the graphene-graphene separation. By interpolating the data, it is obtained that  $F_x\sim\frac{1}{d^{4.2}}$ indicating that the lateral force has a somewhat shorter range coupling compared to $F_0$. 

We can link the $\Delta F_z \sim \beta_d^2$ and $F_x\sim \beta_d$ behavior for $\beta_d<1$ to particular features in Eqs.~\eqref{forcez2}-\eqref{forcex2}. In the case of the vertical force $F_{z}$, the integrand is an even function of the wave vector $q_{x}$. Upon expansion for small $\beta_d$, the linear correction $q_{x}v_{\mathrm{d}}$  renders the integrand odd in $q_{x}$. The subsequent integration over symmetric limits in $q_{x}$ gives a zero contribution, thus the first non-vanishing correction to $F_{z}$ arises only at order $(q_{x}v_{\mathrm{d}})^{2}$, i.e., quadratic in the drift velocity. In contrast, for the lateral force $F_{x}$ its integrand already contains an explicit prefactor $q_{x}$. Expanding the integrand to first order introduces another factor of $q_{x}$, leading to terms of the form $q_{x}^{2}v_{\mathrm{d}}$. This expression survives the integration, yielding a leading-order correction to $F_{x}$ that is linear in $v_{\mathrm{d}}$.

It is also interesting to see how other factors, in addition to the interlayer distance and drift velocity, affect the lateral and vertical force components of the interaction. Fig. \ref{fig4} (a) and (c) shows almost linear dependence upon temperature. The origin of this linear behavior can be traced to the thermal Bose factor entering the fluctuation-induced force. In the relevant low-frequency regime, where $\hbar\omega \ll k_{B}T$, the Bose--Einstein distribution admits the expansion  $\left( \frac{1}{2} +n_B(\omega) \right) \approx \frac{k_B T}{\hbar \omega} + \mathcal{O}\left( \frac{\hbar \omega}{k_B T} \right)$, which shows that $\Delta F_z$ and $F_x$ become proportional to $k_B T/\hbar\omega$ to leading order. Another property that can also be used to modify the interaction is the graphene Fermi energy $E_F$. Increasing $E_F$ enhances the metallic-like nature of the materials leading to stronger Casimir interaction~\cite{Woods2010,Woods2017,Fialkovsky}. This is also observed in the case of current-modified graphene coupling. As shown in Fig. \ref{fig4} (b) and (d), both the lateral and vertical forces experience direct correlation with $E_F$ with a close to linear functional behavior.

\begin{figure}
    \begin{minipage}{0.5\textwidth}
        \centering
    \includegraphics[width=0.9\textwidth]{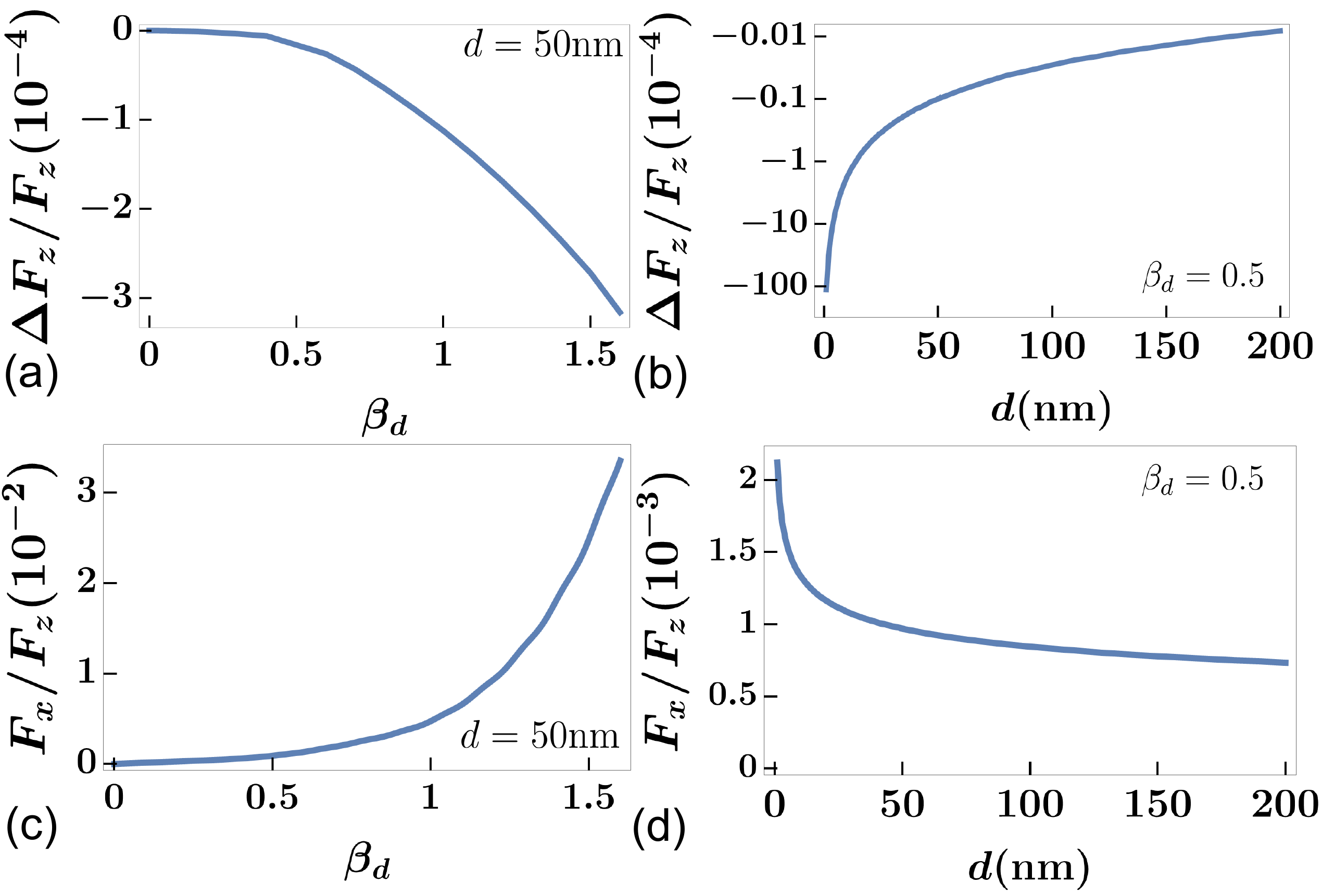}        \caption{Graphene-Graphene Casimir interaction in the presence of drift currents in both layers in opposite directions with the same magnitude: (a) $\Delta F_z/F_z$  as a function of $\beta_d$ at $d=50$ nm; (b) $\Delta F_z/F_z$ as a function of $d$ for $\beta_d=0.5$; (c) $F_x/F_z$ as a function of $\beta_d$ for $d=50$ nm; (d) $F_x/F_z$ as a function of $d$ for $\beta_d=0.5$ Here $\Delta F_z= F_z- F_0$, where $F_z$ is the interaction along the $z$-axis when the drift current flows in the top graphene layer and $F_0=-\frac{3\hbar c\alpha}{32\pi d^4}$ is the equilibrium Casimir force. In all cases, the graphene Fermi level is taken as $E_F=0.1$ eV and $T=300$ K.}\label{fig6}

    \end{minipage}\hfill
\end{figure}

Let us further consider the case of drift currents flowing in both graphene sheets. We take the case of charge carriers flowing with the same velocity, but in opposite directions: carriers in the top layer in Fig. \ref{fig1} flow along the positive $x$-axis with $v_{\rm d}$, while carriers in the bottom layer have $v_{\rm d,2}=-v_{\rm d}$. From the calculations, it is found that the sign of the correction to the forces $F_x$ and $F_z$ are the same as in the previous case with only one layer supporting a current, however the magnitudes are bigger. In particular, for $\beta_d=0.5$, we have $\Delta F_z/F_z (d=1 nm)\approx0.00153$ and $\Delta F_z/F_z (d=50 nm)\approx1.34\times10^{-6}$  for the case of current in one graphene as opposed to $\Delta F_z/F_z (d=1 nm)\approx0.0208$ and $\Delta F_z/F_z (d=50 nm)\approx2.45\times10^{-5}$ for the case of currents in both graphene sheets. Since the distance dependence of $\Delta F_z$ is the same for both single and double-current configurations, their ratio remains essentially independent of separation. The enhancement factor is instead governed primarily by the drift parameter $\beta_d$. The repulsive Casimir force correction along the $z-$axis, given in Fig. \ref{fig6} (a) for the graphene sheets separated at $d=50$ nm when both are carrying currents, still has an approximately quadratic $\beta_d^2$ dependence as in Fig. \ref{fig5}(a). The distance dependence in Fig. \ref{fig6} (b) also shows that the scaling law  remains almost the same compared to the previous case with current in only one layer shown in Fig. \ref{fig5}(b).

Fig.~\ref{fig4} (c) demonstrates that the dependence of the lateral force upon $\beta_d$ experiences nonlinearity even at $\beta_d<1$ unlike the case of $F_x$ for the case of only one graphene layer carrying a drift current. Comparing the magnitudes shows that the lateral force is significantly enhanced for the case of both graphene sheets carying currents. For example,  at $\beta_d=0.5$, $F_x/F_z(d=1 nm)\approx0.00035$ and $F_x/F_z(d=50 nm)\approx0.00016$ for the case of current in one graphene as opposed to $F_x/F_z (d=1 nm)\approx0.00196$ and $F_x/F_z(d=50 nm)\approx0.0009$ for the case of currents in both graphene sheets and the ratio between two cases also remains similar for other distances at each $\beta_d$. 

At the end we discuss the case of currents flowing in both graphene layers but in the same direction. One notes that the two major factors in the Casimir energy in  Eqs.~\eqref{forcez2}-\eqref{forcex2}  come from the Doppler shifted frequencies $\omega'=\omega''=\omega-\boldsymbol{q}\cdot \boldsymbol{v}_d$ and the modified graphene conductivities according to the SFD model in Eq.~\eqref{condcorrection}. We can see that in this case the integrand in Eq.~\eqref{forcex2} is zero, which means that there is no lateral force. An analogous situation may also occur when the two graphene layers are physically moving with the same velocity in same directions~\cite{Volokitin_2001,Intravaia2014,Milton2016}. The non-contact frictional force in this case is zero since there is no relative motion between electrons in the two layers. The correction to the vertical interaction $\Delta F_z$ is also small compared to the results in Figs.\ref{fig5} and~\ref{fig6}. Since the Doppler shifted frequencies are not relevant due to absence of relative motion between the electrons of the two layers, only the SFD modified conductivities are important, therefore the corresponding correction to the equilibrium  vertical component is also less prominent.


Let us further analyze the behavior of the lateral force in the presence of drift currents to highlight the cases discussed earlier. We note that for drift velocities $v_{\rm d,1,2}\ll c$, the Bose distribution factor in Eq.~\eqref{forcex2} can be written as $n(\omega')-n(\omega'')\approx\frac{\hbar}{k_BT}n(\omega)n(-\omega)$. From here, the lateral force can be given as 
\begin{equation}\label{forcex-appr}
F_x=\boldsymbol{\gamma}\cdot ({\bf v}_{d1}-{\bf v}_{d2}) 
\end{equation}
\begin{eqnarray}\label{gamma}
&&\boldsymbol{\gamma}\approx \frac{\hbar^2}{2\pi^3k_BT}\int_{0}^{\infty}d\omega\bigg\{n(\omega)n(-\omega)\int_{q>\omega/c}d^2q \boldsymbol{q}q_x\\
&&\times  e^{-2|k_z|d}\bigg[\frac{{\rm Im}R_{1p}(\omega''){\rm Im}R_{2p}(\omega')}{|1-e^{-2|k_z|d}R_{1p}(\omega'')R_{2p}(\omega')|^2}+(p\leftrightarrow s)\bigg]\bigg\}\nonumber
\end{eqnarray}
In the case of $\beta_{d,1,2}<1$, the Fresnel reflection coefficients can be approximated by Eqs.~\eqref{Rp_expansion}-\eqref{h-expansions} shown in the Appendix~\ref{appendixB}. By expanding the reflection coefficients into the contribution from the Doppler shift in frequency and the modified conductivity of the SFD model the coefficient $\boldsymbol{\gamma}$ can be separated  $\boldsymbol{\gamma}=\boldsymbol{\gamma}_0+\delta\boldsymbol{\gamma}^{Dop}({\bf v}_{d1}-{\bf v}_{d2})/v_F+\delta\boldsymbol{\gamma}^{SFD}({\bf v}_{d1}+{\bf v}_{d2})/v_F$, where $\boldsymbol{\gamma}_0$ accounts for the part given by the equilibrium reflection coefficients. Eq. \eqref{forcex-appr} shows that the lateral force is identically zero for ${\bf v}_{d1}={\bf v}_{d2}$ and it is largest for ${\bf v}_{d1}=-{\bf v}_{d2}$. The drift currents flowing in the opposite directions with the same magnitude maximize the Dopper effect in the coefficient ${\bf \gamma}^{Dop}$ and nullifies the effect of the current induced changes in the graphene conductivity. Eq. \eqref{forcex-appr} with the supporting expressions in Appendix~\ref{appendixB} show that in general the major contribution to the lateral force and frictional coefficient comes from the Doppler frequency shift and the modified SFD conductivity serves as a secondary correction that renormalizes the overall friction coefficient without changing its qualitative dependence on the relative drift velocity.


\section{Conclusion}

In this work, we have investigated fluctuation–induced interactions between two parallel graphene sheets in the presence of steady-state drift currents. By employing the shifted Fermi disk model to capture the non-equilibrium optical response of current-carrying graphene, we resolved the Maxwell stress tensor and derived both the vertical and lateral components of the Casimir force. Our analysis demonstrates that drift currents fundamentally modify the vertical Casimir interaction: while the equilibrium force retains its well-known attractive scaling, the current-induced correction introduces a repulsive contribution. This shorter-ranged repulsion grows sub-quadratically with drift velocity and is most pronounced at small inter-layer separations.

In addition, the presence of drift carriers generates a lateral Casimir-like force, opposing the carrier flow direction and resembling non-contact quantum friction between moving graphene layers. Although considerably smaller in magnitude than the equilibrium vertical force, this lateral component shows linear drift velocity dependence  at small $\beta_d$ and develops nonlinear behavior at larger values, highlighting the rich interplay between charge transport and quantum fluctuations. We further demonstrated how temperature and Fermi energy influence both the vertical and lateral forces, with nearly linear and sublinear dependencies, respectively.

When drift currents are introduced in both graphene layers, the magnitude of the current-induced corrections is further enhanced while preserving their qualitative trends. Taken together, our results establish that current flow provides an effective route to tune Casimir interactions in graphene-based systems, enabling the possibility of engineering fluctuation induced forces at the nanoscale. These findings contribute to the broader understanding of non-equilibrium fluctuation phenomena and open avenues for controlling Casimir forces in two-dimensional materials and van der Waals heterostructures.

\section{Acknowledgment}
This work was supported by the US Department of Energy under Grant No. DE-FG02-
06ER46297. Communications with Prof. Boris Shapiro are also acknowledged.

\appendix
\section{Boundary conditions and Casimir forces}\label{appendix}
In order to calculate the fluctuation-induced forces, we need to first resolve the boundary conditions of the electromagnetic field at the two graphene sheets~\cite{physics5040066,Volokitin_2001}. 
We define the propagation direction of the EM wave to be $\boldsymbol{k}$ and its projection on the horizontal plane to be $\boldsymbol{q}$. The direction $\boldsymbol{n}$ is defined by $\boldsymbol{n}=\boldsymbol{z}\times\boldsymbol{q}$.

We consider the case of a drift current flowing only in the top graphene layer in Fig. \ref{fig1}. The electric field in the space in between the two sheets and outside of the two planes as
\begin{eqnarray}\label{Efield1}
\begin{cases} 
\boldsymbol{E}_1=\boldsymbol{u}_1e^{-i k_z z}&z<0\\\boldsymbol{E}_2=\boldsymbol{u}_2^+e^{i k_z z}+\boldsymbol{u}_2^-e^{-i k_z z}&0<z<d\\
\boldsymbol{E}_3=\boldsymbol{u}_3e^{i k_z z}&z>d
   \end{cases}
\end{eqnarray}
where $\boldsymbol{u}_1, \boldsymbol{u}_2^+, \boldsymbol{u}_2^-, \boldsymbol{u}_3$ are amplitudes for the electric field in the different regions and $k_z=\sqrt{\omega^2/c^2-q^2}$. 

The presence of a drift current in the graphene layer needs specific discussion. The current modifies the graphene conductivity as discussed in the main text. The static current also affects the boundary conditions. However, due to its static nature, this current only affects the boundary condition at zero frequency. We can absorb it as an additional term in $j_{f2}$ at zero frequency and it can be easily seen that this difference in the boundary condition does not contribute to the fluctuation induced forces.

To continue further, we need the boundary conditions for the continuity of the electric and magnetic fields for each layer. For the graphene at $z=0$, these can be written as
\begin{equation}
\boldsymbol{E}_{\boldsymbol{n}}(z=0^+)=\boldsymbol{E}_{\boldsymbol{n}}(z=0^-)
\end{equation}
\begin{equation}
\boldsymbol{E}_{\boldsymbol{q}}(z=0^+)=\boldsymbol{E}_{\boldsymbol{q}}(z=0^-)    
\end{equation}
\begin{equation}
\boldsymbol{z}\times[\boldsymbol{B}_{\boldsymbol{q}}(z=0^+)-\boldsymbol{B}_{\boldsymbol{q}}(z=0^-)]=\frac{4\pi}{c}(\sigma_{1L}\boldsymbol{E}_{\boldsymbol{n}}+\boldsymbol{j}_{f1\boldsymbol q})_{z=0}
\end{equation}
\begin{equation}
-\boldsymbol{z}\times[\boldsymbol{B}_{\boldsymbol{q}}(z=0^+)-\boldsymbol{B}_{\boldsymbol{q}}(z=0^-)]=\frac{4\pi}{c}(\sigma_{1T}\boldsymbol{E}_{\boldsymbol{n}}+\boldsymbol{j}_{f1\boldsymbol n})_{z=0}
\end{equation}
where $\sigma_{L}$ and $\sigma_{T}$ are the longitudinal and transverse graphene conductivities, defined in the main text, and $j_{f1\boldsymbol{q}},j_{f1\boldsymbol{n}}$ are the components of fluctuation induced currents on the graphene layer. Similar relations exist for the graphene layer at $z=d$. From these boundary conditions, we arrive at the following relations for the $\boldsymbol{u}^{\pm}$ in the $0<z<d$ region,
\begin{eqnarray}\label{bound1}
\begin{cases} 
u_{2,\boldsymbol{q}}^++R_{1p} u_{2,\boldsymbol{q}}^-=\frac{k_z}{\frac{\omega''}{2\pi}+k_z\sigma_{1{\rm L}}}j_{f1\boldsymbol{q}}, \\
u_{2,\boldsymbol{q}}^++R_{1s} u_{2,\boldsymbol{n}}^-=-\frac{\omega''}{c^2\frac{k_z}{2\pi}+\omega''\sigma_{1{\rm T}}}j_{f1\boldsymbol{n}},
   \end{cases}
\end{eqnarray}
\begin{eqnarray}\label{bound2}
\begin{cases} 
u_{2,\boldsymbol{q}}^-+R_{2p}e^{2ik_zd}u_{2,\boldsymbol{q}}^+=-\frac{k_z}{\frac{\omega''}{2\pi}+k_z\sigma_{1{\rm L}}}j_{f2\boldsymbol{q}}e^{ik_zd}\\
u_{2,\boldsymbol{n}}^-+R_{2s}e^{2ik_zd}u_{2,\boldsymbol{n}}^+=-\frac{\omega}{c^2\frac{k_z}{2\pi}+\omega''\sigma_{1{\rm T}}}j_{f2\boldsymbol{n}}e^{ik_zd}
   \end{cases}
\end{eqnarray}
where the p-polarized and s-polarized wave propagations are defined in the main text. 

By solving Eq.~\eqref{bound1} together with Eq.~\eqref{bound2} we arrive at the following equations for the fields components
\begin{eqnarray}\label{solnb}
\begin{cases} 
u_{2,\boldsymbol{q}}^+=\frac{k_z}{\Delta_p(\omega',\omega'')}[\frac{R_{1p}}{\frac{\omega'}{2\pi}+k_z\sigma_{2{\rm L}}}e^{ik_zd}j_{f2\boldsymbol{q}}-\frac{1}{\frac{\omega''}{2\pi}+k_z\sigma_{1{\rm L}}}j_{f1\boldsymbol{q}}]\\
u_{2,\boldsymbol{q}}^-=\frac{k_z}{\Delta_p(\omega',\omega'')}[\frac{R_{2p}}{\frac{\omega''}{2\pi}+k_z\sigma_{1{\rm L}}}e^{2ik_zd}j_{f1\boldsymbol{q}}-\frac{e^{ik_zd}}{\frac{\omega'}{2\pi}+k_z\sigma_{2{\rm L}}}j_{f2\boldsymbol{q}}]\\
u_{2,\boldsymbol{n}}^+=\frac{\omega}{\Delta_s(\omega',\omega'')}[\frac{R_{1s}}{c^2\frac{k_z}{2\pi}+\omega'\sigma_{2{\rm T}}}e^{ik_zd}j_{f2\boldsymbol{n}}-\frac{1}{c^2\frac{k_z}{2\pi}+\omega''\sigma_{1{\rm T}}}j_{f1\boldsymbol{n}}]\\
u_{2,\boldsymbol{n}}^-=\frac{\omega}{\Delta_s(\omega',\omega'')}[\frac{R_{1s}}{c^2\frac{k_z}{2\pi}+\omega''\sigma_{1{\rm T}}}e^{2ik_zd}j_{f1\boldsymbol{n}}-\frac{e^{ik_zd}}{c^2\frac{k_z}{2\pi}+\omega'\sigma_{2{\rm T}}}j_{f2\boldsymbol{n}}]
   \end{cases}
\end{eqnarray}
with
\begin{eqnarray}\label{solnc}
\begin{cases} 
u_{2,z}^+=-\frac{q}{k_z}u_{2,\boldsymbol{q}}^+\\
u_{2,z}^-=\frac{q}{k_z}u_{2,\boldsymbol{q}}^-\\
u_{1z}=\frac{q}{k_z}u_{1\boldsymbol{q}}
   \end{cases}
\end{eqnarray}
found from the transversal conditions $\mathbf{k} \cdot \mathbf{E} = 0$ and $\mathbf{k} \cdot \mathbf{B}= 0$. With the above obtained expressions for $\boldsymbol{u}_2^{\pm}$, similar relations for $\boldsymbol{u}_{1,3}$ can be found from the continuity conditions of the electric field $\boldsymbol{E}$.

The field amplitudes obtained in Eq.~\ref{solnb}-\ref{solnc}  are then  substituted into  Eq.~\ref{Efield1} to construct all components of the electric and magnetic fields in the three spatial regions. These explicit field expressions are then used to evaluate the electric- and magnetic-field correlators entering the Casimir force in Eqs. \ref{forcez} and \ref{forcex}, as found from the Maxwell stress tensor. 

As an illustrative example, the term $\langle E_zE_z^*\rangle_{z=0^\pm}$ is calculated explicitly. From the results in Eq.~\ref{solnb}-\ref{solnc} one finds
\begin{eqnarray}
E_z(z=0^+)=E_{2z}(z=0)=\boldsymbol{u}_{2z}^++\boldsymbol{u}_{2z}^-,
\end{eqnarray}
\begin{eqnarray}
E_z(z=0^-)=E_{1z}(z=0)=\boldsymbol{u}_{1z}.
\end{eqnarray}

Then the correlator $\langle E_zE_z^*\rangle_{z=0^+}$ is given by

\begin{eqnarray}
&&\langle E_zE_z^*\rangle_{z=0^+}=\langle( \boldsymbol{u}_{2z}^++\boldsymbol{u}_{2z}^-)(\boldsymbol{u}_{2z}^++\boldsymbol{u}_{2z}^-)^*\rangle\\
&&=\frac{q^2}{|\Delta_p(\omega',\omega'')\bigg|^2}\bigg[\bigg|\frac{R_{1p}+1}{\frac{\omega'}{2\pi}+k_z\sigma_{2{\rm L}}}e^{ik_zd}\bigg|^2\langle j_{f2\boldsymbol{q}}j_{f2\boldsymbol{q}}^*\rangle\nonumber\\
&&+\bigg|\frac{R_{2p}e^{2ik_zd}+1}{\frac{\omega''}{2\pi}+k_z\sigma_{1{\rm L}}}\bigg|^2\langle j_{f1\boldsymbol{q}}j_{f1\boldsymbol{q}}^*\rangle\bigg]\nonumber,
\end{eqnarray} 
Thus the required terms  $\langle E_zE_z^*\rangle_{z=0^+}$ is now expressed in terms of the reflection coefficient, conductivity and the fluctuating currents $\langle j_{f\alpha}(\boldsymbol{q},\omega)j_{f\beta}(\boldsymbol{q},\omega)\rangle=\frac{\hbar \omega}{\pi} \left(\frac{1}{2}+n_B(\omega) \right){\rm Re}[\sigma_{\alpha\beta}(\boldsymbol{q},\omega)]$.  The other terms in Eqs. \ref{forcez} and \ref{forcex} can be obtained in a similar way.

\section{Fresnel reflection coefficients for small drift velocities}\label{appendixB}

In this Appendix we show approximate expressions of the reflection coefficients $R_p,R_s$ within first order in the $\beta_d<$ parameter assuming $v_d<c$, 
\begin{eqnarray}\label{Rp_expansion}
&&R_{p(s)}(\omega - \mathbf{q}\cdot \mathbf{v}_d)\approx R_{p(s)}(\omega) \\
&&\times\Bigg\{ 
1 + \big[ 1 - R_{p(s)}(\omega) \big][ h_{p(s)}^{Doppler}(\omega, \mathbf{q}) +h_{p(s)}^{SFD}(\omega, \mathbf{q}) ]\frac{v_d}{v_F}   \Bigg\},\nonumber 
\end{eqnarray}
\begin{eqnarray}\label{h-expansions}
 &&h_{p(s)}^{Dop}(\omega, \mathbf{q}) =  \left( \pm \frac{q^2}{k_z^2} - \frac{\omega}{g}  \frac{\partial g}{\partial \omega} \right)  \frac{q v_F \cos \theta_{\mathbf{q}}}{\omega}, \\
&& h^{SFD}(\omega, \mathbf{q}) =  \frac{f_{xx (yy)} \cos^2 \theta_{\mathbf{q}} + f_{yy (xx)} \sin^2 \theta_{\mathbf{q}} }{g} .  
\end{eqnarray}
In the above, $R_s(\omega)=\sigma_{{\rm T}}(\omega)/\left(\dfrac{c^2 k_z}{2\pi \omega} + \sigma_{{\rm T}}(\omega) \right)$ and $R_p(\omega)=\sigma_{{\rm L}}(\omega)/\left(\dfrac{\omega}{2\pi k_z} + \sigma_{{\rm L}}(\omega) \right)$  are the Fresnel reflection coefficients as previously defined in the main text. Also, the $+$ sign corresponds to the p-modes and $-$ sign to the s-modes in $h_{p(s)}^{Dop}$.

These approximate expressions enable obtaining Eqs. \eqref{forcex-appr}-\eqref{gamma}, where,\begin{widetext}
\begin{eqnarray}
&&\pmb{\gamma}_0 \approx \frac{\hbar^2}{2\pi^3 k_B T}\int^{\infty}_0 d\omega \bigg\{ n(\omega) n(-\omega) \int_{q>\omega/c} d^2q\mathbf{q} q_x e^{-2|k_z|d} \left[\frac{ \left( {\rm Im}R_{p}(\omega, \sigma_u g) \right)^2}{|1 - e^{-2|k_z|d} R_{p}^2(\omega, \sigma_u g)  |^2} +\left( p \leftrightarrow s \right)\right] \bigg\},
\end{eqnarray}
and

\begin{eqnarray}
&& \delta \pmb{ \gamma}^{Dop (SFD)} \approx \frac{\hbar^2}{2\pi^3 k_B T}\int^{\infty}_0 d\omega \bigg\{ n(\omega) n(-\omega)
   \int_{q>\omega/c} d^2q \; \mathbf{q} q_x e^{-2|k_z|d}  \bigg[
   \dfrac{{\rm Im}\left(R_{p}(\omega, \sigma_u g) \left(1 - R_{p}(\omega, \sigma_u g) \right) h_p^{Dop (SFD)}(\omega,\mathbf{q}) \right)}{{\rm Im}R_{p}(\omega, \sigma_u g) } \nonumber\\
&& \frac{ \left( {\rm Im}R_{p}(\omega, \sigma_u g) \right)^2 }{|1 - e^{-2|k_z|d} R_{p}^2(\omega, \sigma_u g) |^2} {\rm Re} \bigg( \frac{2 e^{-2|k_z|d} R_{p}(\omega, \sigma_u g)^2 (1- R_{p}(\omega, \sigma_u g))}{1 - e^{-2|k_z|d} R_{p}^2(\omega, \sigma_u g) } h_p^{Dop (SFD)}(\omega,\mathbf{q}) \bigg) 
   +\left( p \leftrightarrow s \right) \bigg] \bigg\}.
\end{eqnarray}
\end{widetext}

\bibliography{refs}

\end{document}